\documentclass[twocolumn,10pt]{revtex4-1}

\usepackage{multibib}

\usepackage{cmbright}

\usepackage[english]{babel}

\usepackage{amssymb}
\usepackage{xspace}
\usepackage{amsmath}
\usepackage{graphicx}
\usepackage{floatpag}

\begin{document}

\title{Proteins analysed as virtual knots}

\author{Keith Alexander}
\author{Alexander J Taylor}
\author{Mark R Dennis}

\begin{abstract}
  Long, flexible physical filaments are naturally tangled and knotted, from macroscopic string down to long-chain molecules.  The existence of knotting in a filament naturally affects its configuration and properties, and may be very stable or disappear rapidly under manipulation and interaction.  Knotting has been previously identified in protein backbone chains, for which these mechanical constraints are of fundamental importance to their molecular functionality, despite their being open curves in which the knots are not mathematically well defined; knotting can only be identified by closing the termini of the chain somehow.  We introduce a new method for resolving knotting in open curves using virtual knots, a wider class of topological objects that do not require a classical closure and so naturally capture the topological ambiguity inherent in open curves.  We describe the results of analysing proteins in the Protein Data Bank by this new scheme, recovering and extending previous knotting results, and identifying topological interest in some new cases.  The statistics of virtual knots in protein chains are compared with those of open random walks and Hamiltonian subchains on cubic lattices, identifying a regime of open curves in which the virtual knotting description is likely to be important.
\end{abstract}

\maketitle

\section*{Introduction}
\label{sec:intro}

Proteins are large, complex biomolecules exhibiting folded
conformations, whose precise form and stability are fundamental to
their biological role~\cite{branden98}.  As protein chains can be
thought of as long, tangled curves, it is natural to ask if they can
be \emph{knotted}.  Mathematical knot theory only defines knots in
closed, circular loops~\cite{adams94}, whereas the curves described by
protein chain backbones have distinct endpoints. They are \emph{open
  chains} formed from a string of carbon and nitrogen atoms and may be
`untied' by smooth deformation.  A degree of mathematical compromise
is therefore required to determine whether a given protein chain may
be considered knotted~\cite{millett13,tubiana11}; its termini must
somehow be joined to make a closed curve, without distorting the
protein's configuration.  Various closure constructions have been
proposed~\cite{tubiana11}, generally giving similar results, and
applied to protein chain catalogues~\cite{pdb,knotprot}.  These
investigations have shown that knotting in proteins is in fact very
rare~\cite{knotprot,lua06}, likely owing to the chemical and
mechanical difficulty of forming such structures making them
evolutionarily disadvantageous~\cite{mallamjackson12}.  The
unlikelihood of knotting might suggest an evolutionary advantage when
they do occur~\cite{faisca15,lim2015}, but it remains unclear in most
cases exactly how this manifests~\cite{mallamjackson12,lim2015}.

\begin{figure*}
\includegraphics[width=\textwidth]{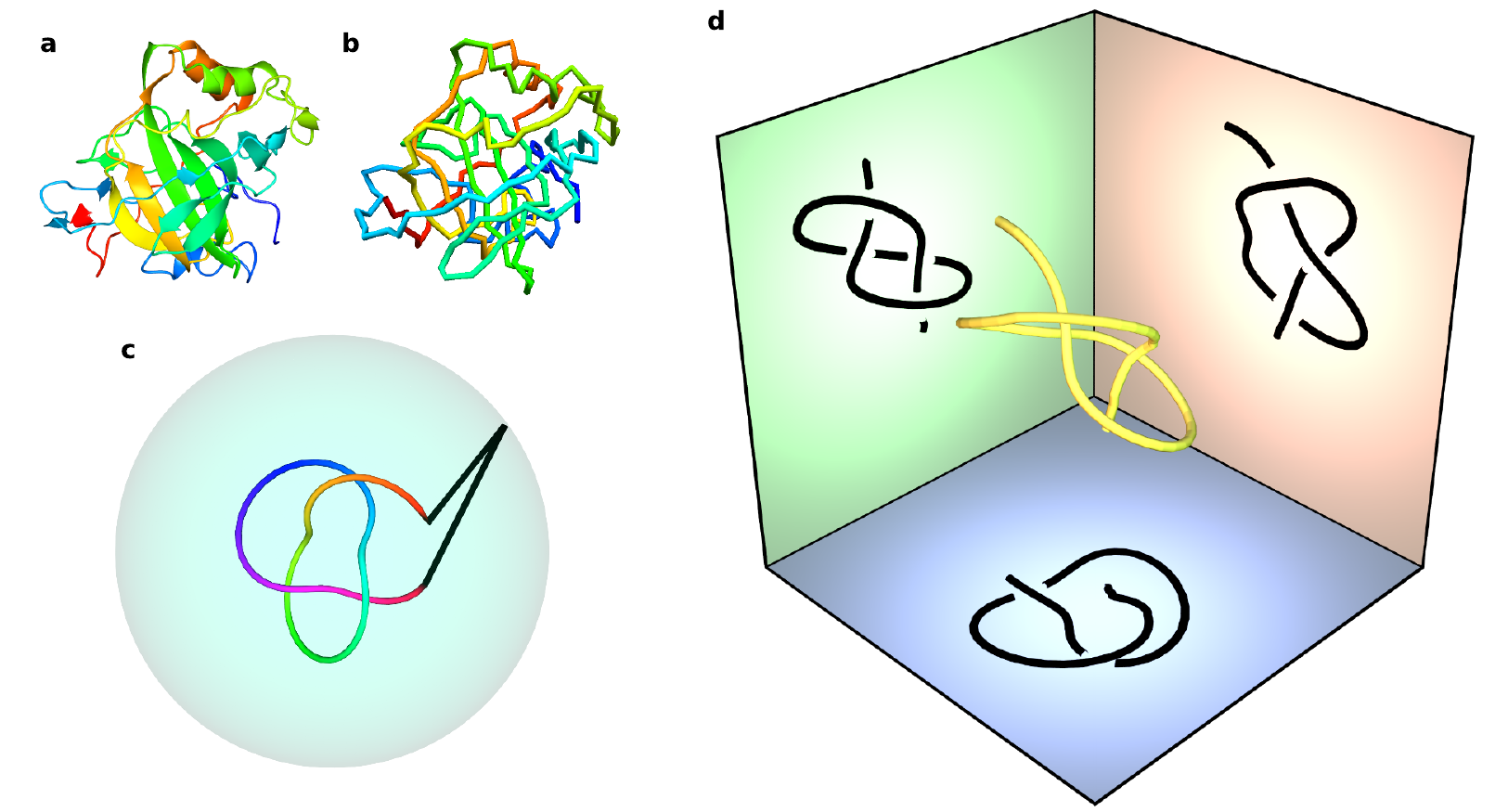}\centering
\caption{ 
  Protein backbone structures as open knotted space curves.  (a) shows the backbone and some secondary structure of the protein with PDB ID 4COQ, chain A (\emph{Thermovibrio ammonificans} alpha-carbonic anhydrase)~\cite{4COQAjames14}, while (b) shows only the backbone chain of carbon alpha atoms as a piecewise-linear space curve. The colouring along the chain serves only to distinguish different regions and does not have physical meaning.  (c) represents the closure of an open curve by straight lines from its termini to a point on a surrounding sphere.  (d) shows a 3-dimensional open curve and its planar projections in three perpendicular directions; each projection here gives an open knot diagram, where each crossing in the projection indicates which strand passes over or under the other.  In this example, each of these projected knot diagrams represents one of two different knot types, as explained in the text.  
}
\label{fig:fig1}
\end{figure*}

Fig.~\ref{fig:fig1}(a) shows a ribbon diagram representation of a
protein chain.
Secondary structures (shown as alpha helices and beta pleated sheets),
as well as bonds other than peptide bonds such as disulphide bonds and
hydrogen bonds, will be ignored in the following analysis, despite
some recent investigation of their conformational
tangling~\cite{haglund14,dabrowski-tumanski16,flapan15,cao05,boutz07,mcdonald93}.
The corresponding protein backbone is shown in Fig.~\ref{fig:fig1}(b)
as a piecewise-linear curve, with each vertex representing a carbon
alpha atom, each connected to its two neighbours, or one neighbour at
the termini.  The most obvious way of closing the backbone into a
closed loop is to join its endpoints with a straight line, but such a
crude procedure usually fails to give a knot representative of the
protein~\cite{tubiana11,millett13}. A method that has become
standard~\cite{millett13,knotprot,lua06} is illustrated in
Fig.~\ref{fig:fig1}(c): straight lines are continued from each
backbone terminus to the same point on a sphere surrounding the curve
(we refer to this as \emph{sphere closure}).  Each point on the
closure sphere gives a closed curve of specific knot type, which may
be the `unknot', equivalent to the trivial circle. Nongeneric closures
where the straight lines intersect the backbone are ignored.  The
sphere is given a large enough radius to avoid small-scale geometrical
effects; in practice, the closing lines can be taken as parallel,
closing `at infinity' (i.e.~the sphere has infinite radius).
Labelling each point on the sphere by the knot resulting from closure
there partitions the sphere surface into `islands' of different knot
types, and the island covering the greatest area may be identified
as the `knot type' of the protein.  The results of the ongoing
\textit{KnotProt} protein survey~\cite{knotprot} (as of Sep 2016)
reveal that according to this definition, 946 of the 159,518 sequence
unique protein chains in the Protein Data Bank~\cite{pdb} (PDB) are
statistically knotted by this measure.

Here we present an alternative analysis of protein knots.  Rather than
\emph{closing} the backbone curve in 3D, we consider the
\emph{projection} of the open curve in every direction.  Each such
projection gives a 2-dimensional open \emph{knot diagram}, a network
of arcs intersecting at \emph{crossing} points, where one arc passes
over the other~\cite{adams94}.  Examples are represented for three
perpendicular projections of a simple open curve in Fig.~\ref{fig:fig1}(d).  The topological analysis is performed on the knot
diagrams by considering them as \emph{virtual knots} via a
\emph{virtual closure} that does not add additional classical
crossings.  Virtual knots are a generalisation of the usual
`classical' knots, that can capture the open nature of the diagram via
new virtual knot types that do not correspond to a closed classical
knot (although classical knots may also result from this
procedure)~\cite{kauffman99}.

The topological character of the open protein backbone chain is fully
characterised by the distribution, over different projection
directions, of different classical and virtual knots resulting from
virtual closure.  An advantage of this new method is that it allows a
more subtle refinement of the knot distribution associated with
an open curve, as the inclusion of virtual knots can better capture
the conformations of backbones where tangling is evident but no single
knot type dominates.  This analysis is particularly suitable for
protein curves, and relates to the distinction between deep knots
(whose knotting is strongly classical) and shallow knots (whose
topological spectrum becomes significantly richer under virtual
knotting).  We quantify these changes, and suggest how these
techniques could apply to specific other systems of open curves.

\section*{Methodology and Results}

\subsection*{Projected open curves and virtual knots}
\label{sec:maths}

In this Section we summarise some basic mathematics of knot and
virtual knot classification~\cite{adams94,kauffman99}. A more complete
summary of both classical and virtual knot theory is given in
Supplementary Note~1.

\begin{figure*}
\includegraphics{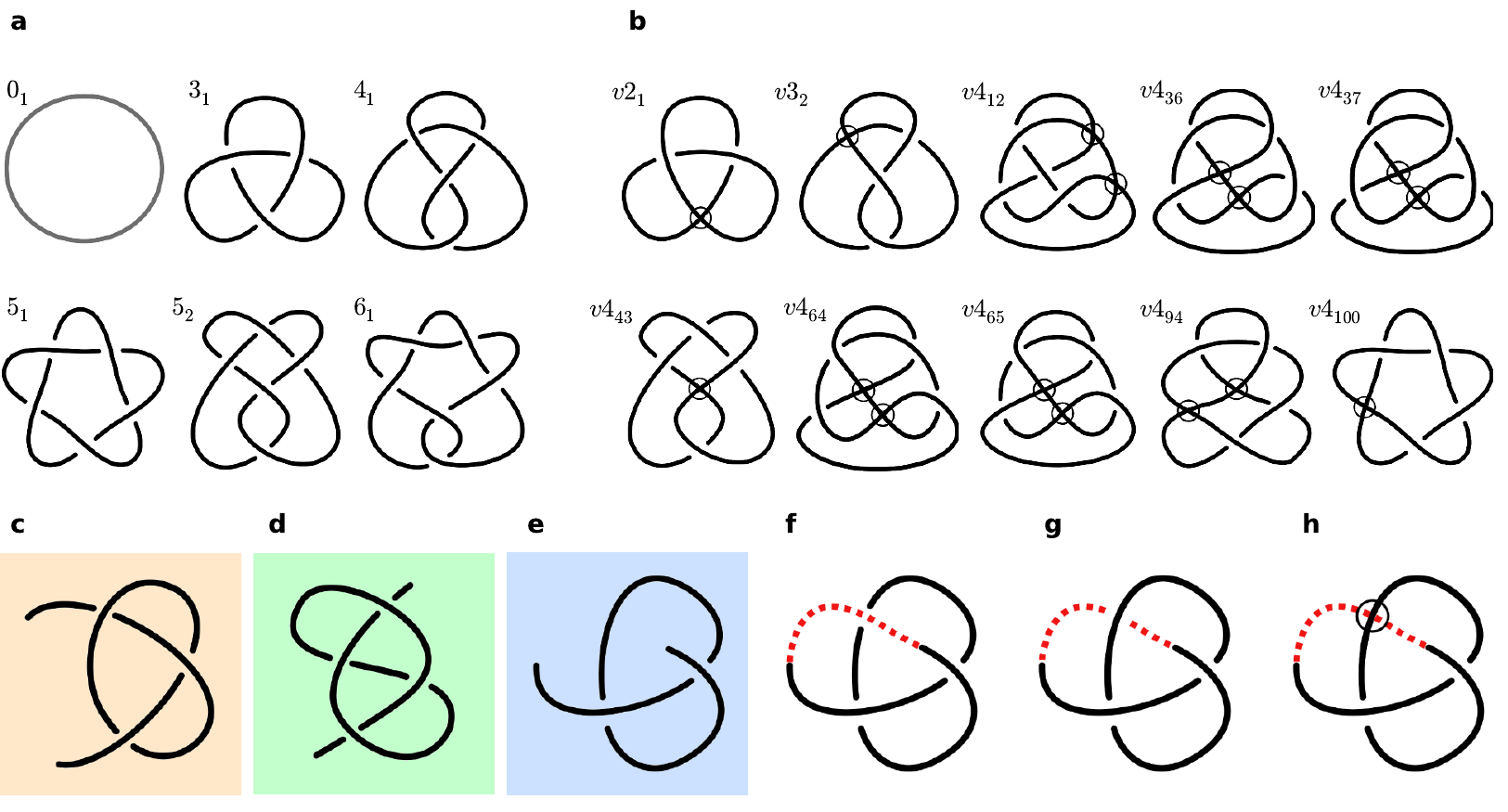}\centering
\caption{ Classical and virtual knot diagrams which could occur as closures of open curves.  (a) shows the first six classical knots in the standard tabulation (including the unknot $0_1$); all but $5_1$ have been identified as dominant knot types in at least one protein~\cite{knotprot}.  (b) shows the virtual knots with $n = 2,3,4$ as tabulated in~\cite{virtualknottable}, all of which can arise as virtual closures of open knot diagrams.  Virtual crossings are shown as circles.  (c)-(h) show examples of open diagrams, which may be identified under virtual closure as classical or virtual knots.  (c)-(e) are equivalent to the projections from Fig.~\ref{fig:fig1}(d). (f) and (g) show (e) closed with a classical arc passing above or below the intervening strands, forming an unknot $0_1$ and trefoil knot $3_1$ respectively, while (h) shows (e) closed instead with a virtual crossing to produce the knot $v2_1$.  }
\label{fig:fig2-1}
\end{figure*}

Knots are labelled and ordered in \emph{knot tables}
\cite{rolfsen76,hoste98,knotatlas,knotinfo} according to their
\emph{minimal crossing number} $n$, which is the minimum number of
crossings a 2-dimensional diagram of the knot may have~\cite{adams94}.
The closed knots are labelled $n_m$, where $m$ counts the knots of
minimal crossing number $n$, not distinguishing enantiomeric pairs
with opposite chirality (indeed, we do not distinguish between such
pairs here, although it would be possible to do so).  Examples of some
simple knots appear in Fig.~\ref{fig:fig2-1}(a), such as the trefoil
knot $3_1$ (the only knot with $n = 3$) and the only two five-crossing
knots $5_1$, $5_2$.  Composite knots, in which more than one knot is
tied in a single curve, do not appear in protein
chains~\cite{knotprot}.  A given knot has many possible conformations,
which may have arbitrarily many crossings in projection; equivalent
conformations (which can be deformed into one another without cutting
and joining) are said to be \emph{ambient isotopic}, and their
diagrams can be related by a sequence of \emph{Reidemeister moves}
\cite{adams94} (see Supplementary Fig.~1).

Open curves are technically not knots, as they do not form a closed
loop and so have endpoints. We instead close the endpoints with an arc
that makes \emph{virtual crossings} with the other arcs, which do not
distinguish over or under crossing.  Under this \emph{virtual closure}
each open diagram represents a \emph{virtual knot}~\cite{kauffman99},
a generalisation of normal knot diagrams.  All the topological
information is contained within the classical crossings (in this
sense, the virtual crossings represent `not closing' the curve), so
the virtual crossings capture the ambiguities between the different
classical closures. A given open knot diagram has the same virtual
knot type under all possible virtual closures, although this may still
represent a classical knot (and all classical knots can arise from
virtual closure).  This procedure is illustrated in
Fig.~\ref{fig:fig2-1}(c)-(e): in (c) and (d) the endpoints can be
closed with no additional virtual crossings, in both cases
representing the classical trefoil knot $3_1$, while in (e) there is
no way to avoid crossing an intervening strand.
Fig.~\ref{fig:fig2-1}(f) and (g) show the ambiguity of classical
closure, resulting in the unknot $0_1$ and trefoil knot $3_1$
respectively, while in (h) the virtual closure produces a single
virtual knot.  We note that open knot diagrams could instead be
considered in the slightly wider class of \emph{classical
  knotoids}~\cite{turaev12}, whose isotopies are determined by
augmented Reidemeister moves which forbid endpoints from passing
over/under any strand of the curve, but although knotoids form their
own topological classes~\cite{turaev12,gugumcu16} they have not yet
been robustly tabulated (see Supplementary Note 1). Our method
corresponds to the virtual closure of the classical
knotoid~\cite{gugumcu16}.

\begin{figure*}
\includegraphics{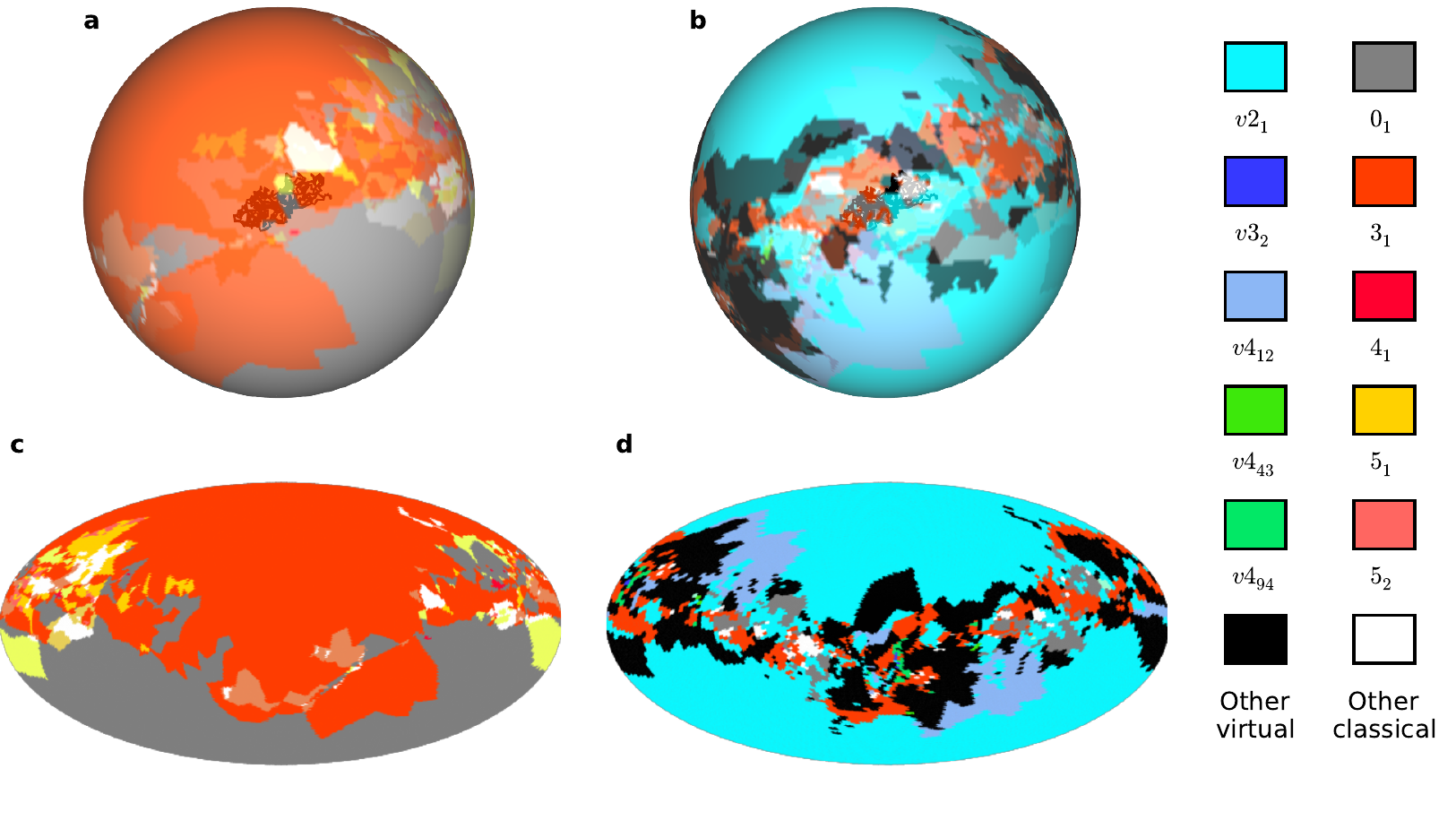}\centering 
\caption{ Classical and virtual knot types found amongst different projection/closure directions for a protein backbone chain. The protein backbone shown has PDB ID: 4K0B, chain A (\emph{Sulfolobus solfataricus} S-adenosylmethionine synthetase)~\cite{4K0BAwang2014}.  The points of each sphere are coloured according to the knot type (classical or virtual) found by closure/projection in that direction. (a) shows the classical knots resulting from sphere closure at each point, with knot types according to the legend, while (c) shows the same colouring in an area-preserving (Mollweide) projection of the sphere, making its entire area visible. (b) shows the virtual knot types resulting from projection in each direction on the sphere, and (d) the same colouring again by projection. These images are constructed from sampling $10,000$ directions in each case.  Antipodal points on the sphere are always associated with the same knot type under virtual closure (up to possibly distinct mirrors for certain virtual knot types), but may produce different classical knots on sphere closure.  }
\label{fig:fig2-2}
\end{figure*}

Tabulations of virtual knots~\cite{virtualknottable,kauffman99} follow
the same ordering logic.  We denote virtual knots with a prefix `$v$',
i.e.~$vn_m$ where $n$ is again the minimum classical crossing number
(there is no relationship between the classical $n_m$ and virtual
$vn_m$), with examples given in Fig.~\ref{fig:fig2-1}(b).  $n$ is
invariant to (appropriately generalised) ambient isotopy and `virtual'
Reidemeister moves (see Supplementary Note 1).  As with the classical
tabulation, all mirror-symmetric partners are considered to be
equivalent.  Not all virtual knots can arise from virtual closure of
open diagrams.  The only ones that can occur are those that can be
drawn with all virtual crossings adjacent, with no classical crossings
in between (i.e.~along the closure arc); the examples with up to 4
classical crossings are shown in Fig.~\ref{fig:fig2-1}(b). There are
still many more of these than classical knots for given $n$: the
classical (virtual) count is 1 (0) for $n = 0$; 0 (1) for $n = 2$; 1
(1) for $n = 3$; 1 (8) for $n = 4$, etc.

In practice, the knot type (classical or virtual) of each closed
diagram is found through calculation of \emph{knot
  invariants}~\cite{adams94,rolfsen76,kauffman99,virtualknottable},
which are functions of the diagram whose values depend only on its
(classical or virtual) knot type.  Most readily-calculated invariants
fail to distinguish certain distinct knots~\cite{adams94}, so we
identify types by the characteristic signatures of a set of
invariants, calculated sequentially until the knot type is clear
(after additionally simplifying each diagram algorithmically using
Reidemeister moves).  It is more computationally efficient to
calculate polynomial invariants at specific values rather than
symbolically, and we consider them at certain roots of
unity~\cite{taylor16}.  For classical knots, invariants are: the
\emph{Alexander polynomial}~\cite{adams94} $\Delta(t)$ at $t = -1$
(the knot determinant~\cite{adams94}), $t = e^{2\pi i/3}$ and
$t = -i$.
For virtual knots we use the \emph{generalised Alexander polynomial}~\cite{kauffman03,virtualknottable} $\Delta_g(s, t)$ at $(s,t) = (-1, e^{2\pi i/3})$, $(-1, i)$, $(e^{2\pi i/3}, i)$; and the \emph{Jones polynomial} $V(q)$~\cite{jones85,adams94,rolfsen76,kauffman1987} at $q = -1$.
Classical knots have $\Delta_g = 0$.

We will present results on knotting in terms of the fractions of
directions giving different knot types under sphere or virtual
closure.  Fig.~\ref{fig:fig2-2}(a)-(b) demonstrate this structure for
an example protein chain, by colouring the sphere according to the
knot types found in each direction from both of the closure methods,
while (c) and (d) show the same results in an (area-preserving)
Mollweide projection of the sphere area such that its entire surface
is visible; this projection is preferred in later figures.  In the
sphere closure map (c), many of points are unknotted (grey), yet 59\%
give a trefoil knot $3_1$, which therefore dominates and so this
backbone was determined by \cite{knotprot} to be $3_1$ knotted.  The
smaller islands where closures form more complex knots make up less
than 7\% of the sphere area.  In the corresponding virtual closure map
(d), the virtual knot $v2_1$ is associated with much of the area
identified as $0_1$ or $3_1$ in (c), now appearing in 54\% of
different projections.  This curve therefore has strong virtual
character, and its virtual knot type reflects the ambiguity of the
open curve between the unknot and trefoil knot.

\subsection*{Analysis of the Protein Data Bank}
\label{sec:results}

We now present the results of our survey of knotting in the Protein
Data Bank (PDB)~\cite{pdb}, using both sphere closure and virtual
closure.  Following the methodology of the KnotProt
database~\cite{knotprot}, we constructed a minimal set of distinct
chains from the 121,532 structures recorded in the PDB, analysing only
each sequence unique chain in a given protein and rejecting chains
containing artefacts.  We additionally restrict attention to chains
that have not been made obsolete by more recent measurements.  The PDB
records for some of the remaining proteins have broken chains (where
the chain conformation is uncertain), which we close with straight
lines.  This gives a total of 159,518 protein chains for analysis.
For each chain, we close/project to 100 different points on the sphere
(approximately uniformly distributed following the method
of~\cite{rakhmanov94}), considered sufficient for reasonable numerical
confidence at acceptable computational cost~\cite{millett13}.

The sphere closure analysis of KnotProt found 946 knotted chains,
including 871 trefoil ($3_1$) knots, 45 occurrences of $4_1$, 27 of
$5_2$ and 3 of $6_1$ (at time of comparison: Sep 16). Our
corresponding analysis gives instead $972$ knotted chains, including
894 of $3_1$, 48 of $4_1$, 27 of $5_2$ and 3 of $6_1$, but does
include all but one of the KnotProt-identified chains, leaving 27
additional knot detections.  These discrepancies appear to arise from
small differences in methodology, particularly in rare occasions where
very severe chain breaks are present; 17 of our extra detections are
considered knotted by one or both of the alternative protein knots
databases pKNOT~\cite{lai07}, or Protein Knots~\cite{kolesov07}.  We
therefore consider that our sphere closure methodology accurately
detects protein knotting for the purpose of comparison with virtual
closure.

In the above results, the knot associated with an open chain is the
most common single knot type occurring over sphere closure in
different directions (i.e.~the modal average).  Although this
methodology is natural, this can miss certain interesting cases; for
instance, a chain closing in different directions to 40\% unknot, 30\%
$3_1$ and 30\% $4_1$ would be considered unknotted, despite giving
some knot for the majority of closure directions. Such cases are much
more frequent under virtual closure, as many more knot types are
possible, and the resulting maps are correspondingly more complex as
shown in Fig.~\ref{fig:fig2-2}.  We therefore introduce new classes of
knotting associated with open chains, based on the definition that an
open chain is \emph{unknotted} only if it appears to be $0_1$ in over
50\% of closure directions; it is otherwise considered knotted, in
some sense. For sphere closure, if a single (nontrivial) knot type
occurs in at least 50\% of directions we call this \emph{strongly
  knotted}, while if the sum of different nontrivial knot types occurs
for at least 50\% of directions, but no single type does, we call this
\emph{weakly knotted}.  This does not significantly affect the 972
protein knots discussed above; almost all (968) are strongly knotted
by these definitions, with 7 further chains being weakly knotted.  The
choice of threshold at 50\% is somewhat arbitrary, and the number of
curves identified as unknotted rises (falls) as it is increased
(decreased).

\begin{figure*}
\floatpagestyle{empty}
\includegraphics{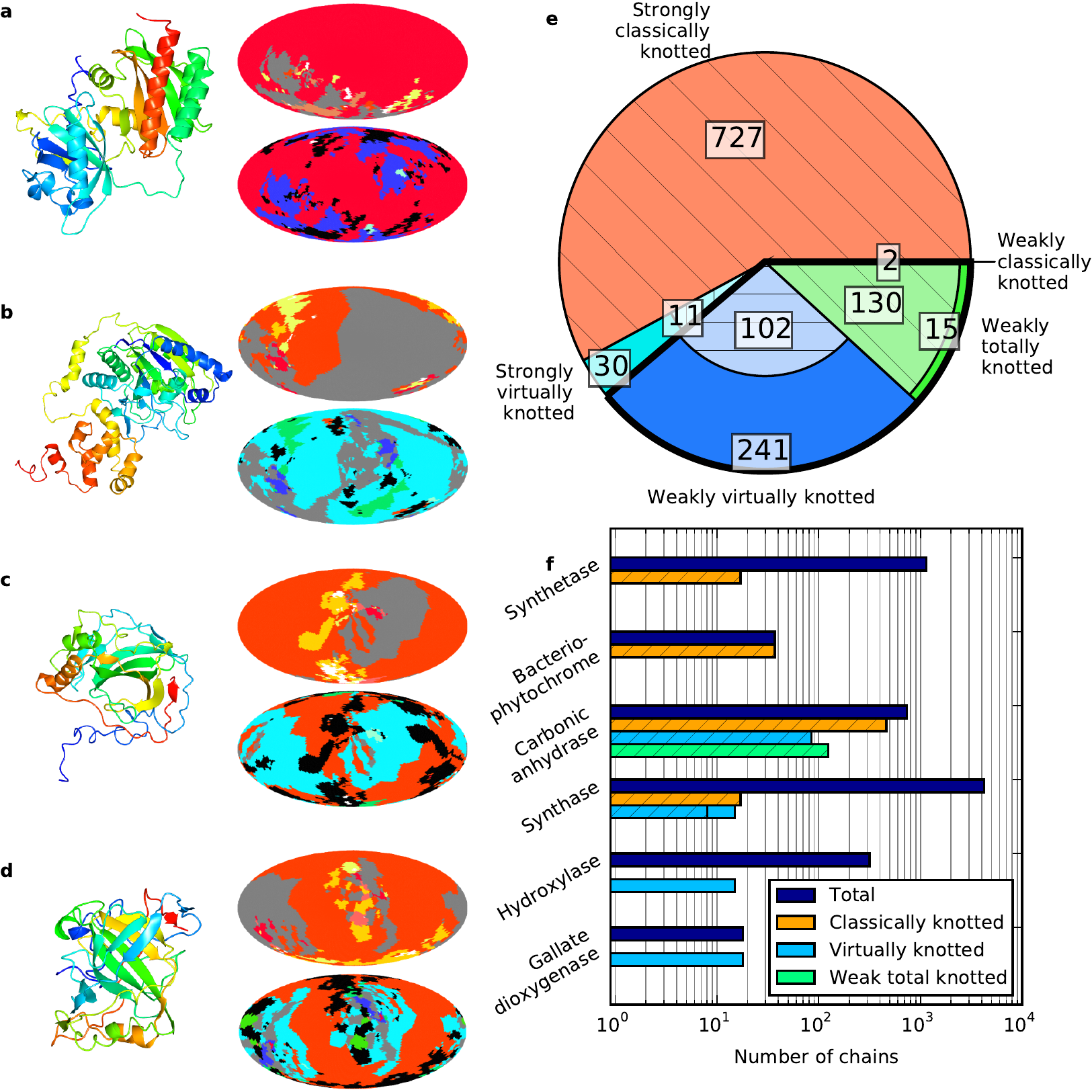}
\caption{ Results of virtual closure analysis for knotting in the Protein Data Bank.  Knot categorisations follow the main text; strong classical (virtual) knotting where more than 50\% of projections form the same classical (virtual) knot type; weak classical (virtual) knotting when over 50\% of projections form classical (virtual) knots but no single knot type dominates, and weak total knotting where the unknotting fraction does not exceed 50\% but no other specific class dominates.  (a)-(d) give examples of knot type maps (see Fig.~\ref{fig:fig2-1}) for protein chains in these different classes, with colours following the legend of Fig.~\ref{fig:fig2-2}.  The upper map in each case shows the results of sphere closure, while the lower shows virtual closure: in (a) PDB ID: 4E04, chain A (\emph{Rhodopseudomonas palustris} RpBphP2 chromophore-binding domain)~\cite{4E04Abellini2012}, which is classically knotted in both cases, although the virtual closure reveals new structure; in (b) PDB ID: 3WKU, chain B (\emph{sphinogobium sp. SYK-6} extradiol dioxygenase) ~\cite{3WKUBsugimoto2014}, which is not knotted under sphere closure but is strongly virtually knotted under virtual closure; in (c) PDB ID: 4XIX, chain A (\emph{Chlamydomonas reinhardtii} carbonic anhydrase)~\cite{4UELAel2010}, which is knotted under both sphere and virtual closure, weakly virtually knotted in the latter; and in (d) PDB ID: 3KIG, chain A (\emph{Homo sapiens} carbonic anhydrase II mutant)~\cite{3KIGwischeler2011}, which is knotted under sphere closure and exhibits weak total knotting on virtual closure.  (e) summarises the number of protein chains in each knotting class under virtual closure.  (f) shows knot types found amongst selected categories of protein chain names, and their distribution amongst knotting classes.  In both (e) and (f), hatched areas represent chains which were also identified as knotted under sphere closure.  }
\label{fig:fig3_1}
\end{figure*}

Under virtual closure, the different projections may include a mixture
of virtual and classical knot types. We refine the distinction of
strong and weak knotting to distinguish some major categories of knot
character, calling a chain \emph{strongly classically (virtually)
  knotted} where a single classical or virtual knot type appears in
more than 50\% of virtual closures from different projection
directions (e.g. strongly trefoil knotted or strongly $v3_2$
knotted). A chain is instead \emph{weakly classically (virtually)
  knotted} if no knot type is so individually common, but a
combination of different classical (virtual) knot types alone
contributes to over 50\% of projection directions (e.g. 30\% $v3_1$,
30\% $v2_1$ and 40\% $0_1$ is weakly virtually knotted). In all
other cases, no specific classification dominates, and we call the
curve \emph{weakly totally knotted}. All of these weak classes
represent knots with significant topological character that is not
consolidated in forming a single deep knot. Examples of protein chains
according to these classifications are shown in
Fig.~\ref{fig:fig3_1}(a)-(d), and the identifications may vary
significantly from the results obtained by sphere closure: (a) is
strongly classically knotted according to both analyses; (b) was
unknotted on sphere closure but is strongly virtually ($v2_1$)
knotted on virtual closure; (c) was strongly $3_1$ knotted on sphere
closure but is weakly virtually knotted on virtual closure; and (d)
was strongly $3_1$ knotted on sphere closure but on virtual closure is
weakly totally knotted.

Altogether we find 1258 protein chains falling into one of these
topological classes, 283 more than in our sphere closure analysis.
The mix of their different classifications is summarised in
Fig.~\ref{fig:fig3_1}(e). As with the sphere closure analysis, most
of these protein chains are strongly classically knotted (727 cases,
all of which were also strongly classically knotted under sphere
closure, and mostly the knot $3_1$), and weak classical
knotting is still negligible (2 cases, 7 under sphere closure).
Strong virtual knotting is much less common, occurring in 41 cases, 30
previously considered unknotted under sphere closure. These are
cases where two classical knot types compete with comparable area
contribution under sphere closure, and in all but one case the
competition is between $0_1$ and $3_1$; the virtual knots are
therefore strongly $v2_1$ knotted (the remaining example is
$v4_{43}$ between classical types $0_1$ and $5_2$).

The remaining protein chains are weakly knotted in some form; 343 are
weakly virtually knotted (around a third of which were not
topologically interesting under sphere closure), and 145 are weakly
totally knotted (most of which were dominated by a classical knot
under sphere closure). The new detections here represent curves that
cannot be easily identified with a single classical knot type because
their conformations are similar to multiple classical knots. This is
demonstrated in Fig.~\ref{fig:fig3_1}(c), whose knot types under
sphere closure suggest little of the complexity evident in its virtual
closure map; this feature is typical of the weak virtual knots, which
for this reason include most of the new chains that appeared unknotted
under sphere closure. These knots may be interpreted as being rather
shallow, as small modifications to the chain can relatively
significantly affect the maps. The weakly totally knotted chains are
similar but with the classical knots a little deeper in the chain, as
in the example of Fig.~\ref{fig:fig3_1}(d), where the clarity of the
chain's trefoil knot character is muted but not removed under virtual
closure.

These various classifications of strong and weak knots form a loose
way of capturing the forms of knotting and tangling exhibited in
protein backbone curves, with physical implications for the depth of
the knots in the chain. The distribution of these classes is uneven
amongst the protein chains; for instance, all 46 examples of $4_1$
under sphere closure remain strongly $4_1$ knotted under virtual
closure, suggesting consistently small virtual character.  Knotting is
also not equidistributed amongst different protein classes:
Fig.~\ref{fig:fig3_1}(f) shows a breakdown of the the different
classes of knotted open chain by protein chain name, for families in
which knotting has previously been observed to
cluster~\cite{knotprot}, as well as families where new virtual
character appears.  Virtual knotting appears significant amongst
carbonic anhydrases, in which the knots are known to be rather
shallow, and all knots found under virtual closure also appear under
sphere closure. In contrast, the virtual knots amongst synthases are
almost all newly identified, with previously discovered strong
classical knots being deep enough to remain unchanged by the analysis.
Further, the families of hydroxylases and gallate dioygenases contain
several examples of virtual knotting, and neither family showed any
evidence of knotting under sphere closure. It is unsurprising that the
levels of topological complexity are consistent among members of the
same protein families,
 as they arise from consistent features in their
secondary and tertiary structures, 
but it is important that virtual knotting has its
own distribution among protein chain names, distinct from that of
classical knotting.

\subsection*{Comparison with random open chain ensembles}
\label{sec:randomwalks}

The virtual closure technique for describing knotting is applicable to
any open space curve, but the the presence of virtual knots relies on
particular geometric characteristics of the curve. It is unclear if
proteins express these in a generic fashion, or if virtual knotting is
a particularly good (or bad) descriptor of their backbone chains, and
for comparison we perform a preliminary analysis by sphere closure and
virtual closure for other families of random open curves.  These are
drawn from two statistical ensembles: open random walks, and open
subchains of Hamiltonian walks on a cubic lattice. In order to
investigate the new information provided by virtual knotting we use a
simplification of the scheme in the previous section, considering an
open curve as `knotted' if over 50\% of directions yield a knot on
sphere closure (i.e. either strong or weak classical knotting), and
`virtually knotted' if over 50\% of projection directions are
virtually knotted (i.e. either strong or weak virtual
knotting). Virtual closure is a useful technique for ensembles where
the virtual knotting probabilities are comparable to or higher than
closure knotting probabilities, otherwise most curves will take the
same strong knot type under both analyses.  The main parameter against
which knotting is compared is \emph{closing distance fraction}
(CDF)---the distance between the curve's endpoints divided by its
total length---which varies from 0 for a closed loop, to 1 for a
straight line.

Random walks consist of a sequence of random linear steps, whose
limiting, long-length statistical behaviour is that of Brownian
motion.  Their geometry and topology is quite well understood; for
sufficiently long walks, the statistics are independent of the
specific model, tending towards the characteristic
Brownian fractal behaviour~\cite{falconer97}.  The probability of
knotting in closed random walks has been well
investigated~\cite{orlandini07}.  Random walks tend not to be a good
model for proteins, but nevertheless are good models for other
physical systems~\cite{flory53,orlandini07,taylor16}, and are a
convenient comparison model for knotting of open chains in the absence
of physical constraints.

\begin{figure*}
\includegraphics{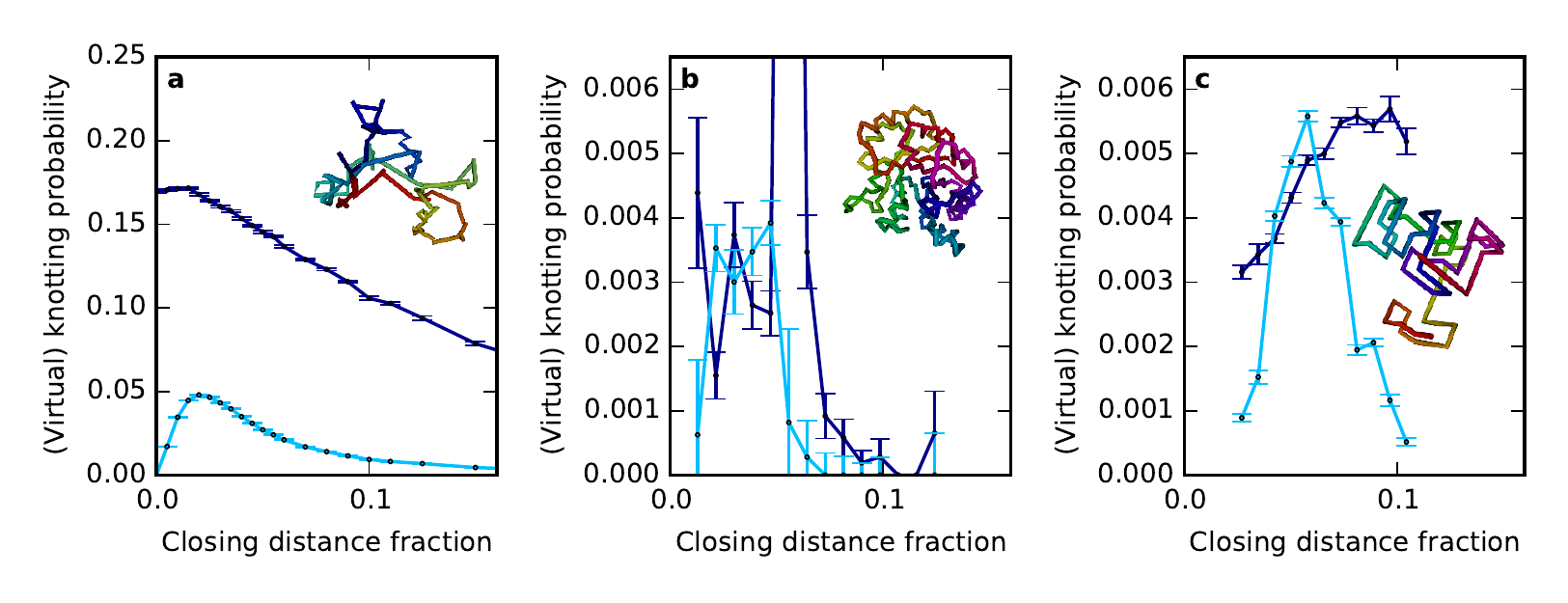}\centering
\caption{ Knotting and virtual knotting probabilities in different open curve ensembles.  The closing distance fraction (CDF) is the ratio of the distance between the open curve's endpoints with respect to the total curve length.  Knotting probabilities are given (a) $6\times10^6$ open random walks of length 100; (b) all 159,518 proteins analysed in the previous Section, with various lengths and binned according to CDF; (c) $5.5\times10^6$ length-75 subchains of Hamiltonian walks on cubic lattices of side length 6, binned by CDF.  In (c), the large fluctuations reflect correlations implicit in the lattice.  In each figure, the inset shows a typical example of the curve ensemble, coloured red to blue by hue along its length to distinguish different regions of the curve.  Error bars represent the standard error on the mean probability of the knot statistic.  }
\label{fig:rws}
\end{figure*}

Fig.~\ref{fig:rws}(a) shows the statistics of knotting upon sphere and
virtual closure for a set of random walks with 100 steps generated via
the method of~\cite{cantarella12}, with inset showing a sample random
walk.  The advantage of this particular ensemble is that the CDF can
be directly controlled, but for all distances knotting is
significantly more common than virtual knotting (this is most probable
around a CDF of 0.025, where about 5\% of the random walks are
virtually knotted, but even at this value classical knotting is at
least 3.5 times as common).  This qualitative result appears to hold
for random walks of very different lengths (not shown).  These results
are not surprising as knots in random walks can easily be small,
localised deep within the chain.

This contrasts strongly with the equivalent results for proteins,
shown in Fig.~\ref{fig:rws}(b), which combine all protein chains from
the previous Section despite their backbones being of many different
lengths (from tens to thousands of angstroms and up
to
$\sim$3300 carbon atoms in the backbone chain).  The comparatively
small number of protein chains mean the statistics are only useful for
qualitative comparison.  Nevertheless, virtual knotting appears far
more likely relative to classical knotting, possibly becoming more
dominant around a CDF of 0.025.

Unlike random walks, protein backbones are characterised by relatively
compact geometries (such as the inset to Fig.~\ref{fig:rws}(b)), and
aspects of this this can be reproduced by simple mathematical models
of random chains.  In Fig.~\ref{fig:rws}(c), we give the results for
one such model: a subchain of a Hamiltonian walk~\cite{lua06}, that
is, a path on a cubic lattice of fixed size, visiting every vertex
once and every edge no more than once.  Such curves form a confined,
folded structure due to the strict boundaries of the finite lattice.
The geometry and topology of proteins are best approximated when the
Hamiltonian segment is much shorter than this, such that the lattice
confinement is not strong, and random lattice walks of this type can be
efficiently generated up to lattice side lengths of at least
10~\cite{lua04}.

Fig.~\ref{fig:rws}(c) shows the knotting and virtual knotting
sampled from $5.5\times10^6$ random Hamiltonian subchains with length
75 on a cubic lattice of side length 6,
with these parameters
chosen to approximate the knotting probabilities in
Fig.~\ref{fig:rws}(b).  Here the virtual knotting is strong relative
to closure knotting, comparable to proteins but very unlike random
walks, and the probability of virtual knotting exceeds that of
classical knotting across the small range $0.04 \lesssim$ CDF
$\lesssim 0.055$.  This trend appears to be highly robust to different
parameters; even if the lattice is saturated, such that knots are very
common, virtual knotting exceeds classical knotting over approximately
the same range.  These results emphasise that virtual knotting is a
generic feature of certain geometrical classes of curves, arising from
relatively weak geometric constraints even in the absence of the
physical complexity of protein chains.

\section*{Discussion}
\label{sec:discussion}

We have shown that the backbones of protein chains, as well as other
open curves, can be described topologically in terms of virtual
knotting.  Through the method of virtual closure of projections, open
chains are found to have a much wider set of topological classes than
the classical knots in closed curves, and we have found many examples
of different virtual knot types, in projections of protein chains.
Nevertheless, virtual knotting dominates relatively few proteins, and
the virtual knot types which do occur are only a small fraction of the
possible virtual knots.  In some cases this can be thought of as
representing a more nuanced characterisation of `almost' knotted
curves, softening the binary distinction between knotting and
unknotting imposed by traditional closure methods.  In the analysis of
proteins the most dominant virtual class is the weak virtual knots,
where no single knot type is most prevalent but less than 50\% of
projected diagram directions are unknotted. These curves are the most
topologically ambiguous, and cannot be associated with a definite knot
type.

Protein chains express several geometrical properties that might be
expected to encourage virtual knotting: as they fold they curve and
twist into relatively small, chemically bound structures such that
their projections have many crossings; the endpoints of the protein
backbone are often within or near the surface of the structure, such
that projections in different directions produce distinctly different
knot diagrams; and the physical limits on their curvature and overall
tangling mean that knots are rarely unambiguous local structures but
inherently involve the entire protein chain.  This is not true for
random walks, and indeed virtual knotting was found to be less
significant in them, although Hamiltonian subchains, which do have
some of these properties, were found to be particularly strongly
virtually knotted.  We expect that virtual knotting analysis will be
most relevant in other systems of open curves with compact
configurations.  A mechanism that might encourage virtual knots in
physical systems is tight confinement, such as that of a curve
confined within a sphere (e.g.~DNA within a viral
capsid~\cite{marenduzzo2013topological,diao14}) but also between less
confining barriers such as adjacent
planes~\cite{orlandini13,micheletti12}, which might privilege certain
projection directions.

Under the virtual closure analysis, a single chain can project to many
different classical and virtual knot types, which we have summarised
by emphasising `strong' dominating single knot types, or the `weak'
classes of mixed classical and virtual knot types.  Although
this captures some differences in the tangling of open curves, it
ignores the rich structure of knot types in the projected map, other
details of which may be necessary to understanding the 3D spatial
conformation of the open chain.  Including virtual knots may be
important to understand these maps, not only because the number of
possible types is increased, but also because they generally occur in
between classical knot types (seen clearly in Figs~\ref{fig:fig2-2} and~\ref{fig:fig3_1}(b)-(d)), even in chains which
are mostly unknotted.  This extra discriminatory ability would be
useful in any classification of open curve geometry according to these
deeper projection correlations, and could also apply to any
investigation of topological character over time in dynamic systems,
capturing the intermediate stages between unknotting and unambiguous
classical knotting.

Although we have focused our discussion on the statistics of virtual
knotting in protein backbone chains, the analysis only requires that
the curves are open-ended; virtual closure is a refinement rather than
an alternative to existing methods of analysing knotting in open
curves, and can be applied anywhere in place of sphere closure of the
open chain.  This could include other aspects of protein knotting,
such as slipknotting in which knots appear in subsections of
the curve before disappearing as the rest of the curve `unthreads'
itself~\cite{sulkowska12}.  Many examples of slipknotting have been
found in proteins~\cite{knotprot}, and tracking the knot type across
subchains of the full protein backbone produces a slipknotting
  fingerprint.  Extending these methods to include virtual knots via
virtual closure would be natural, as virtual knots would typically
occur at transitions between different classical knot types.  The
methodology also can be further extended, for example to systems of
multiple open curves under similar average closures which extend in
the same fashion to the theory of virtual links (and potentially to a
wider class of virtual knot types), and may even extend to other
knot-like objects such as protein lassos~\cite{dabrowski-tumanski16}.

\section*{Methods}
\label{sec:methods}

\textbf{Knot detection by sphere closure of open curves.}  For each
open chain (here, a protein backbone or random walk), each direction
(point on a sphere around the curve) is associated with a type of
knot. For the sphere closure analysis, the endpoints of the open curve
are closed by extending them `to infinity' in this direction, giving a
closed curve of a specific classical knot type. In practice, the 3D
chain is projected in the plane perpendicular to this direction, then
the diagram closed with a straight line that passes \emph{over} every
intervening arc of the diagram. Each open curve is projected and
analysed in 100 approximately uniformly distributed closure
directions, chosen using the algorithm of~\cite{rakhmanov94}. Previous
work has verified that 100 closure directions is usually sufficient to
determine the significant statistical behaviour of closures in
different directions~\cite{millett13}, and so alternative
approximately-uniform samplings should reproduce the same
statistics. For each projection, the resulting knot diagram is
algorithmically simplified using Reidemeister moves (see Supplementary
Note 1), then the knot type identified through the calculation of knot
invariants as described in the main text. The invariant used is the
modulus of the Alexander polynomial, $|\Delta(t)|$, evaluated at each
of $t=-1$, $t=e^{2\pi i/3}$ and $t=i$, computed using a standard
scheme~\cite{orlandini07}. The Alexander polnomial is used because it
can be calculated in polynomial time in the number of crossings of a
knot diagram (more discriminatory invariants are harder to calculate),
but it is still sufficient to distinguish unambiguously knots with up
to at least 8 crossings; more complex knots may have invariants taking
the same values, but these complex conformations are rare and never
dominate in protein chains (for instance, the next knot with the same
Alexander polynomial as the trefoil knot $3_1$ has 13 crossings, and
no simpler knot agrees at the roots of unity we consider either). For
simple knots this choice of three evaluation values is just as
discriminatory as the full Alexander polynomial, but more convenient
for numerical calculation.

\textbf{Knot detection by virtual closure of open curves.}  For the
virtual closure analysis of open curves, the selection of projection
directions proceeds according to the above method, but the projected
diagram in a given direction is closed instead with virtual crossings
and simplified algorithmically using both classical and virtual
Reidemeister moves (see Supplementary Note 1). The same 100 projection
directions are used (and 100 directions appear sufficient to
distinguish knot types as in the sphere closure analysis). Virtual
knots require different invariants, we use the generalised Alexander
polynomial $\Delta_g(s, t)$ at certain pairs of arguments ($s=-1$,
$t=e^{2\pi i/3})$), ($s=-1$, $t=i$) and ($s=e^{2\pi i/3}$,
$t=i$). Unlike the classical knots, even the simple virtual knots
$v2_1$, $v3_1$ and $v4_{94}$ have equal
$\Delta_g(s,t)=(-s^{-2} + s^{-1})t^2 + (s^{-2} - 1) t^{-1} + (-s^{-1}
+ 1)$.
In these cases we additionally calculate the Jones polynomial $V(q)$
at $q=-1$~\cite{adams94}, which requires exponential time in the
crossing number but unambiguously distinguishes all these
examples. Some more complex virtual knots would also be ambiguous to
these measurements but, as with the classical knots in sphere closure,
are far more complex than those appearing in protein chain
closures. Some virtually closed diagrams represent classical knots, in
which case $\Delta_g(s,t)=0$ and the Alexander polynomial is used as
above. These cases are still occasionally complex virtual knots with
vanishing $\Delta_g$, so we further calculate whether the classical
knots produced from over- and under- closure of the virtual crossing
arc are the same; although not proven, we anticipate that if their
knot types differ the diagram likely represents a virtual knot, whose
type we do not identify. In practice, such cases make up a negligible
fraction of total projections and do not limit the analysis.

\textbf{Numerical analysis of protein backbone chains.} The protein
chains are obtained from the list of all recorded protein molecules in
the Worldwide Protein Data Bank (PDB)~\cite{berman03}. In each case
the .pdb protein record is downloaded and parsed using
ProDy~\cite{prody}.  In particular, we parse the atomic coordinates of
each carbon alpha atom, and reconstruct the protein backbone by
connecting these sequentially with straight lines.  This is an
approximation to the true NCCNCC backbone. In some cases there are
missing residues in the PDB record, and here the distant carbon alphas
across any breaks are connected with straight lines to create one,
continuous open curve for each protein chain.  We also ignore
heteroatom structures.  Where protein
chain names are referenced in the text, these are as recorded in the
PDB. 
Protein ribbon structure images were
created using CCP4mg~\cite{mcnicholas2011}.

\subsection*{Acknowledgments}

The authors are grateful to Ben Bode, Paula Booth, Neslihan
G\"ug\"umc\"u, Lou Kauffman, Annela Seddon, Joanna Sulkowska and Stu
Whittington for valuable discussions. This research was funded by the
Leverhulme Trust Research Programme Grant No. RP2013-K-009, SPOCK:
Scientific Properties of Complex Knots. Keith Alexander was funded by
the Engineering and Physical Sciences Research Council. This work was
carried out using the computational facilities of the Advanced
Computing Research Centre, University of Bristol.

\subsection*{Author contributions}

KA carried out the protein analysis and virtual knotting
routines. AJT carried out the classical knot identification
and random chain analysis, and suggested the original
problem. MRD directed the study and drafted the manuscript.

\subsection*{Competing financial interests}

The authors declare no competing financial interests.

\appendix

\renewcommand\tablename{Supplementary Table}
\renewcommand\figurename{Supplementary Figure}
\setcounter{figure}{0}

\begin{figure*}[h]
  \includegraphics[width=\textwidth]{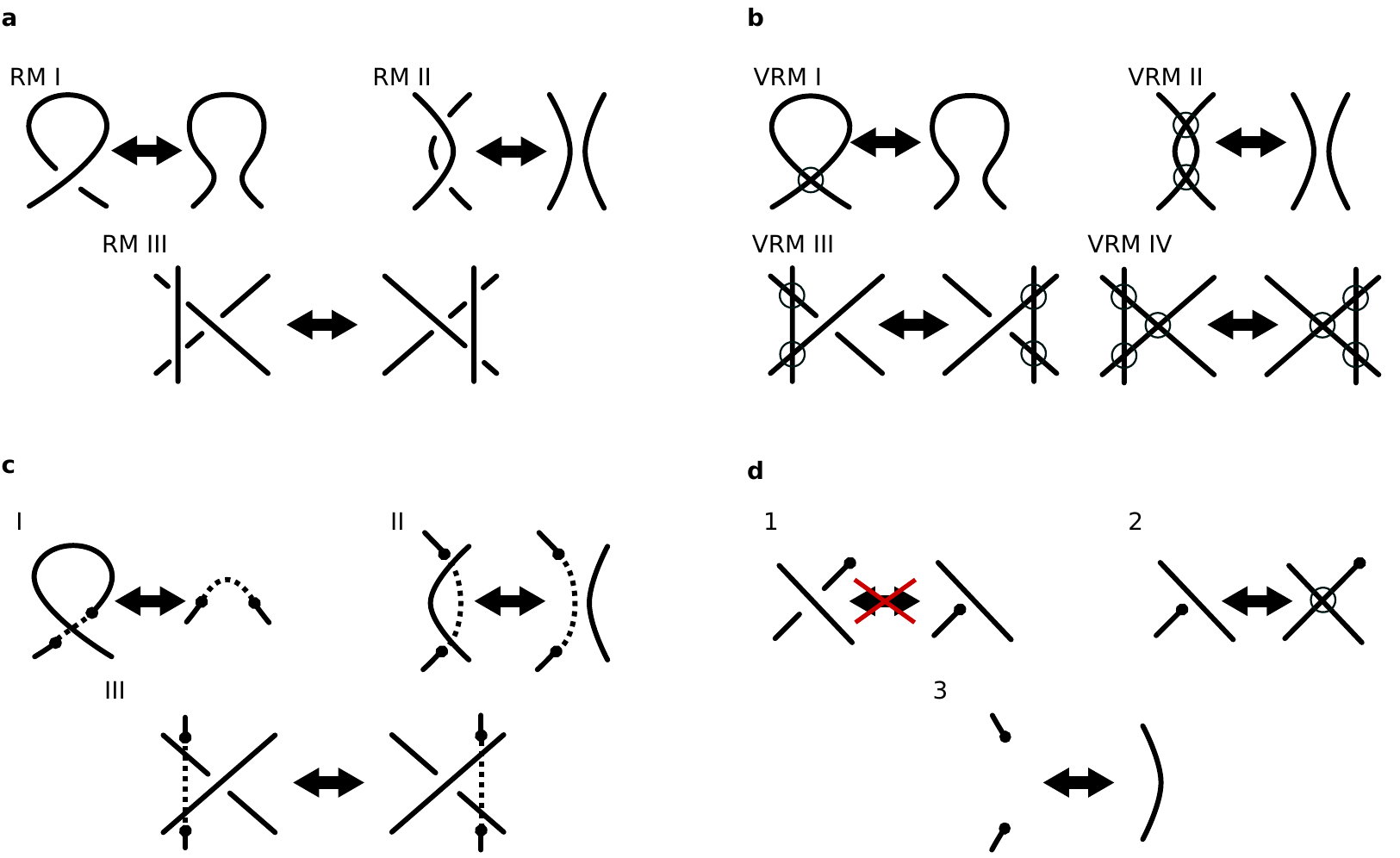}
  \caption{Classical and virtual Reidemeister moves, and other
    algorithmic operations on open knot diagrams. The Reidemeister
    moves are local modifications to adjacent crossings of knot
    diagrams that do not change their topology. For each move, the
    rest of the curve (not shown) is assumed not to interact with the
    depicted region. A suitable combination of Reidemeister moves can
    (alongside planar isotopies) transform a given knot diagram to any
    other representing the same knot. (a) shows the classical
    Reidemeister moves involving only classical crossings, with
    standard labellings. (b) shows the virtual Reidemeister moves
    (virtual crossings are circled), which can involve changes in both
    classical and virtual crossings. 
    (c) shows analogues of the virtual Reidemeister moves as closures
    of open curve diagrams with endpoints, in which case the
    Reidemeister changes clearly do not affect the topology resulting
    from virtual closure of the open strand.  (d) highlights other
    relations that can be applied to open curve diagrams, whose
    application does not affect the virtual knot type resulting from
    virtual closure (disallowing moves 2 and 3 here reproduces the
    knot diagram relations of classical knotoids~\cite{gugumcu16SI}). In
    (c) and (d), each endpoint of the open endpoints of the open curve
    is marked by a black circle. }
  \label{fig:supp_fig_1}
\end{figure*}

\begin{figure*}[h]
\centering
  \includegraphics{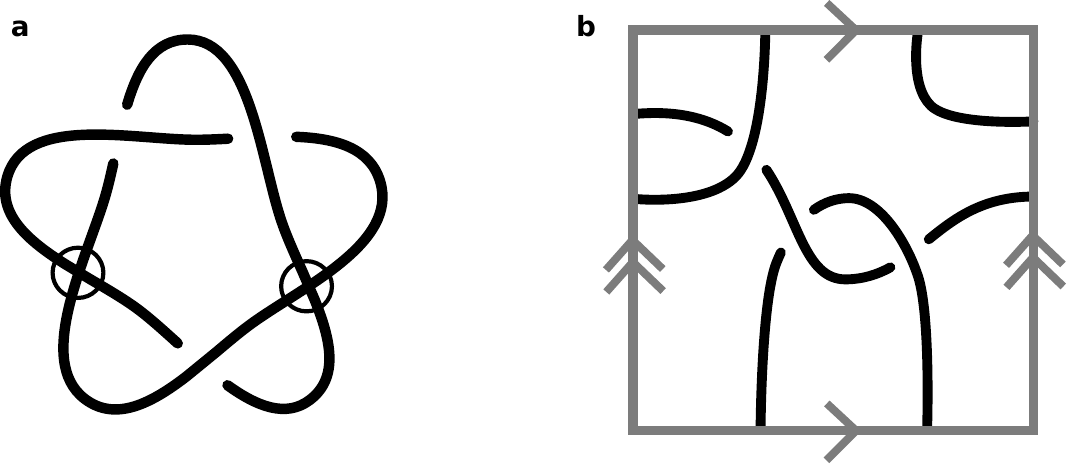}  
  \caption{Virtual knot $v3_7$, a virtual knot which cannot be formed from the closure of a projected open curve. The usual virtual knot diagram is shown in (a) while the presentation in (b) is depicted on the surface of a torus.}
  \label{fig:supp_fig_2}
\end{figure*}

\begin{figure*}[h]
  \includegraphics[width=\textwidth]{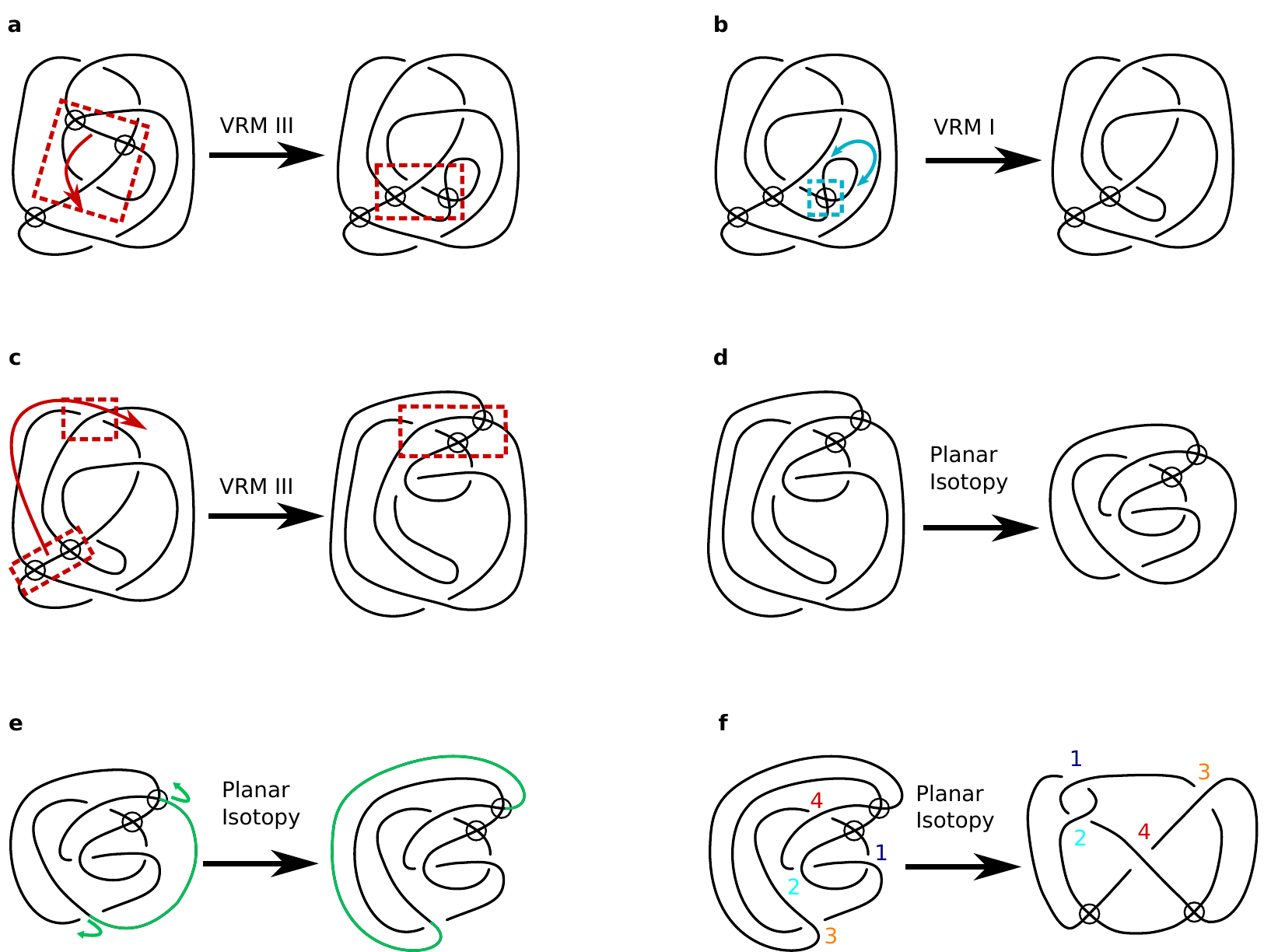} 
  \caption{Transformation between two depictions of the virtual knot
    $v4_{64}$. The initial conformation in (a) is that depicted as
    $v4_{64}$ in the virtual knot table
    of~\cite{virtualknottableSI}. This conformation could not arise from
    virtual closure of an open curve, as its virtual crossings do not
    lie sequentially along a single arc. An alternative presentation
    of this knot from the genus one table of~\cite{andreevna14SI}
    (labelled there as $4_8$), is shown in (f), and does have such a
    conformation, although it is difficult to see by eye that this is
    the same knot as $v4_{64}$. (b)-(e) show how (a) may be
    transformed to (f) by a combination of virtual Reidemeister moves
    and planar isotopies of the knot. In (e), the planar isotopy
    moving the green strand across the knot is not directly allowed by
    the virtual Reidemeister moves, but as the knot diagram is
    implicitly drawn on $\mathbb{S}^2$ this represents the strand
    passing `behind' the sphere (or on the plane, passing through
    infinity). In general it is difficult to test whether two
    (virtual) knot diagrams can be related this way, hence the
    calculation of knot invariants which remove the need for
    diagrammatic manipulation.}
  \label{fig:supp_fig_3}
\end{figure*}

\clearpage

\begin{turnpage}
\begin{table}[h]
  \begin{center}
    {\renewcommand{\arraystretch}{2}
    \begin{tabular}{| c | c | c | c | c | c | c | c | c |}
    \hline
    Knot & $\Delta(-1)$ & $\Delta(e^{2\pi i/3})$ & $\Delta(i)$ & $\Delta_g(-1, e^{2\pi i/3})$ & $\Delta_g(-1, i)$ & $\Delta_g(e^{2\pi i/3}, i)$ & $V(-1)$ & $\Delta_g(s, t)$ \\ \hline
    $0_1$ & 1 & 1 & 1 & 0 & 0 & 0 & 1 & 0 \\ \hline
    $3_1$ & 3 & 2 & 1 & 0 & 0 & 0 & 3 & 0 \\ \hline
    $4_1$ & 5 & 4 & 3 & 0 & 0 & 0 & 5 & 0 \\ \hline
    $5_1$ & 5 & 1 & 1 & 0 & 0 & 0 & 5 & 0 \\ \hline
    $5_2$ & 7 & 5 & 3 & 0 & 0 & 0 & 6 & 0 \\ \hline
    $6_1$ & 9 & 7 & 4 & 0 & 0 & 0 & 9 & 0 \\ \hline
    $v2_1$ & - & - & - & 3 & 4 & 5 & 2 & $s^2 + s^2/t + st - s/t - t - 1$ \\ \hline
    $v3_2$ & - & - & - & 3 & 4 & 5 & 4 & $s + s/t + t - 1/t - t/s - 1/s$ \\ \hline
    $v4_{12}$ & - & - & - & 3 & 8 & 5 & 5 & \parbox{8cm}{\centering $s^2/t + s^2/t^2 - st - s/t - 2s/t^2 - t^2 + t - 1/t + $ \\  $ 1/t^2 + 2t^2 / s + t/s + 1/{st} - t^2/s^2 - t/s^2$} \\ \hline
    $v4_{36}$ & - & - & - & 3 & 8 & 9 & 4 & \parbox{8cm}{\centering $t - 1 - 1/t + 1/t^2 + t^2/s - 2t/s + 2/{st} - 1/{s t^2} - $ \\  $ t^2/s^2 + t/s^2 + 1/s^2 - 1/{s^2t}$} \\ \hline
    $v4_{37}$ & - & - & - & 0 & 4 & 8 & 2 & $s^4 - s^4/t^2 + s^3t + s^3/t^2 - s^2t + s^2/t - st^2 - s/t + t^2 - 1 $ \\ \hline
    $v4_{43}$ & - & - & - & 7 & 8 & 9 & 5 & $s^3 + s^3/t + s^2t - s^2/t - st - s$ \\ \hline
    $v4_{64}$ & - & - & - & 3 & 4 & 0 & 4 & $s^2/t + s^2/t^2 - s/t - s/t^2 + t^2/s + t/s - t^2/s^2 - t/s^2$ \\ \hline
    $v4_{65}$ & - & - & - & 3 & 8 & 9 & 5 & \parbox{8cm}{\centering$-s^2t + s^2 + s^2/t - s^2/t^2 - st^2 + 2st - $ \\  $ 2s/t + s/t^2 + t^2 - t - 1 + 1/t$} \\ \hline
    $v4_{94}$ & - & - & - & 3 & 4 & 5 & 5 & $s^3 + s^3/t + s^2t - s^2/t - st - s$ \\ \hline
    $v4_{100}$ & - & - & - & 3 & 0 & 8 & 4 & $s^4 + s^4/t + s^3t - s^3/t - s^2t + s^2/t + st - s/t - t - 1$ \\ \hline
    \end{tabular}
    }
    \caption{Table of numerical knot invariants for each knot shown in
      Fig.~2(a) and 2(b) of the main text. Included are $\Delta(t)$, the
      Alexander polynomial at the numerical values used, $\Delta_g(s, t)$ the generalised Alexander
      polynomial, both at the numerical values used and the full symbolic expression, and $V(q)$ the Jones polynomial. Virtual
      knots have no Alexander polynomial and so these columns are
      omitted. In the cases where chiral mirrors give different knots,
      only one mirror is given.}
    \label{table:supp_tab_1}
  \end{center}
\end{table}
\end{turnpage}

\clearpage

\subsection*{Supplementary Note 1: Topological Background} \label{sec:supp_note_1}

\subsubsection*{Classical Knot Theory}

In this Supplementary Note we summarise some extended details of mathematical knot theory as used in deriving the results of the main
text. 
Further details can be found in standard elementary texts~\cite{adams94SI,rolfsen76SI,sossinsky02SI}.

Classical knot theory deals with embeddings of the circle, $S^1$ (i.e.~closed, non-intersecting curves), in three-dimensional space $\mathbb{R}^3$. 
Any given embedding has a distinct \emph{knot type}, which is invariant under ambient isotopies (it may change only when the curve passes through itself).
It is usual to represent knots using a 2-dimensional planar \emph{knot diagram}, which can be thought of as a plane projection of the three-dimensional space curve, annotated with the extra information of which strand passes over the other at each self-intersection of the diagram (called a \emph{crossing}).
All the information about the three-dimensional knot type is contained in such a diagram, and smooth deformations (i.e.~ambient isotopies) of the three-dimensional space curve lead to smooth isotopies of the knot diagram, which may change the configuration of the crossings. 
In two-dimensional knot diagrams, the changes are represented by combinations of \emph{Reidemeister moves} as in Supplementary Fig.~\ref{fig:supp_fig_1}(a); applying these local moves in conjunction with \emph{planar isotopies} of the knot diagram can transform between any two diagrammatic representations of the same knot, and equivalently any ambient isotopy of a closed three-dimensional space curve is corresponds to a combination of planar isotopies and Reidemeister moves in any projection of the knot. 

The standard tabulations of knots, in \emph{knot tables} as discussed in the main text, are ordered according to their \emph{minimal crossing number} $n$ -- the smallest number of crossings a diagram of the knot can have.
For instance, the trivial circle can be projected to a plane without self intersection (i.e.~no crossings), and so has minimal crossing number $n = 0$ and is labelled $0_1$. 
There are no knots with $n = 2$, and one with $n = 3$, the trefoil knot, denoted $3_1$.
The labelling $n_m$ continues, where $m$ is an arbitrary index amongst knots with the same $n$. 
These labels are standard, following original tabulations up to $n=10$ published over 100 years ago, with more recent extensions using consistent indices~\cite{rolfsen76SI,knotatlasSI,knotinfoSI}. 
Some simple knots from these tabulations are shown in Fig.~2(a).
$0_1$ (the \emph{unknot}), then $3_1$, $4_1$, etc. 
The knots appearing in knot tables are \emph{prime knots}; \emph{composite knots}, made up of two or more prime knots tied in the same curve, are also possible and are tabulated according to the composition of their prime
factors~\cite{adams94SI}. 
All the tools of knot theory apply equally to composite knots, but they do not occur significantly in any known protein chain, and are not considered further here.

It is natural to follow the curve of a knot, which endows an \emph{orientation} to the knot (choosing an orientation is an arbitrary choice that does not affect the results of topological calculations).  Observing the relative orientation of the strands at a crossing determines the \emph{sign} of the crossing, either positive or negative.  A crossing has the same sign even if the curve's orientation is reversed.  The minimal diagram of a figure-8 knot $4_1$ has two positively signed crossings and two negatively signed, and in fact is isotopic to its mirror image.  On the other hand, all three crossings of the minimal trefoil knot $3_1$ have the same sign, and are all reversed on its mirror image.  Knots such as the trefoil are thus \emph{chiral knots}, and this chirality not directly represented in the tabulation (i.e.~there are two enantiomeric trefoil knots which cannot be be smoothly deformed into one another).  Other chiral knots are $5_1$, $5_2$ and $6_1$ in Fig.~2(a) of the main text; the others are achiral.  We do not distinguish between chiral knot pairs in our analysis, although knot invariant quantities such as used to distinguish knots below could be used to do so.

In practice the knot type of a space curve is determined as follows.
First the curve is projected to a 2D knot diagram, which contains all the topological information in its ordered set of signed crossings along the curve. 
Several topological notations representing this information are standard~\cite{adams94SI,rolfsen76SI}; we use below the \emph{Gauss code}, constructed from an arbitrary starting point and orientation for the curve.
As each new crossing is encountered along the curve, it is labelled $1,2,\ldots$ in order as it is encountered. 
The Gauss code is the ordered list of these crossing numbers as they occur along the curve, together with whether the curve
passes over or under the intersecting strand, represented by using a positive number in the former case and negative in the latter (this is not the same as the crossing sign); each crossing must be encountered exactly twice before reaching the original starting point, once positive and once negative. 
For instance, a Gauss code for a minimal diagram of the trefoil knot $3_1$ is $1, {-2}, 3, {-1}, 2, {-3}$, and for a minimal figure-8 knot $4_1$ is $1, {-2}, 3, {-1}, 4, {-3}, 2, {-4}$. 
It is obvious that changing the starting point on the curve cyclically permutes the crossings encountered, but all the Gauss codes obtained this way, or by changing numeric labels (as long as each crossing retains a unique label) represent the same knot diagram. 
The Gauss code written in this way also does not specify the chirality of the original three-dimensional curve, this information is contained in the local twisting of the two strands around one another and is sometimes included in extended Gauss code notations.
Crossings which can be removed by Reidemeister moves I and II can be easily identified in a Gauss code; if crossing $k$ occurs adjacent to itself, $\pm k, \mp k$ then it can be removed by Reidemeister move I, and if $\pm k, \pm k+1, \ldots \mp k, \mp k+1$ (or $\mp k+1, \mp k$), then crossings $k, k+1$ can be removed by Reidemeister move II.

All knot diagrams can be represented by Gauss codes, but in fact not all Gauss code sequences represent knot diagrams; for instance, the sequence $1, -2, -1, 2$ appears to be a consistent Gauss code of only two crossings, which cannot be simplified by Reidemeister moves, and no knot has $n = 2$. 
On attempting to draw a diagram with this code, one finds it would be necessary for there to be one extra crossing to allow the curve to return to its starting point.
In fact, this is the Gauss code of the open diagram shown in Fig.~2 (e) of the main text, and Gauss codes for open diagrams, and their relation to virtual knots, is the subject of the next section.

It can be practically difficult to calculate the knot type of a diagram coming from a projection of a complicated 3D space curve, which may have many more crossings than its minimal number $n$.
These crossings would represent local geometrical or biochemical features that do not affect the overall knot type; the knot diagrams found from closures of protein backbones often contain several hundred crossings.
Our knot identification proceeds first by algorithmic simplification via removal of crossings, repeatedly applying Reidemeister moves I and II
where they would remove crossings locally (Supplementary
Fig.~\ref{fig:supp_fig_1}(a)), as discussed above.
There is no known efficient method to produce minimal knot diagrams in this way as Reidemeister move III may also be essential to simplify the diagram but does not directly reduce the crossing number. In the case of protein backbones, this occasionally produces minimal diagrams but in most cases tens to hundreds of crossings remain.

The knot types of the simplified diagrams are calculated using \emph{knot invariants}, quantities that depend only on the knot type but are calculated from the geometrical information of the curve, i.e.~they can be calculated from only the information in a Gauss code and their value is invariant to Reidemeister moves. 
Much of mathematical knot theory is devoted to the study of knot invariants, and many types are known. 
For instance, the minimal crossing number discussed above is a knot invariant~\cite{adams94SI}, but there is no simple algorithm to calculate it directly from a presentation of a knot. 
The minimal crossing number also demonstrates that most invariants do not perfectly distinguish knots~\cite{adams94SI}, as multiple different knots can clearly have the same number of crossings in their minimal projections; for instance, both $5_1$ and $5_2$ in Fig.~2(a) have $n=5$. More discriminatory invariants exist but are generally relatively difficult to calculate.

For knot identification we use knot invariants that can be calculated efficiently (ideally in low order polynomial time in the number of crossings), while still discriminating knots sufficiently well. 
In particular, we choose invariants which leave no ambiguity between the knots common on closure of proteins such as those in Fig.~2(a) of the
main text.  
Some protein closures produce complex knots whose knot type cannot be uniquely identified using these efficient invariants, but these occur only rarely and do not impact our analysis.  
For classical knots, we employ only the \emph{Alexander polynomial} $\Delta(t)$, which can be found as the determinant of a matrix whose
rows and columns relate to the crossings of a projected diagram and can be easily constructed from a Gauss code~\cite{orlandini07SI}. 
Computing symbolic matrices numerically is relatively slow, and we instead use the values of $|\Delta(t)|$ evaluated at roots of unity
$t=-1$, $t=\exp(2\pi i/3)$ and $t=i$, such that the calculation can be performed using floating point arithmetic (this does not introduce
appreciable error). 
Each of these is individually a lesser knot invariant, but together they have discriminatory power comparable to the full Alexander polynomial up to at least 11 minimal crossings (certainly sufficient for the relatively simple knots that appear in protein chains).

Many knot invariants, including the Alexander polynomial, are available from standard online resources including the Knot Atlas~\cite{knotatlasSI} for all knots with up to 15 crossings, and KnotInfo~\cite{knotinfoSI} for a wider selection of invariants up to 12 crossings. 
Supplementary Table~\ref{table:supp_tab_1} shows values of $\Delta(t)$ at the roots of unity used above, for each of the simple knots that appear most commonly in protein chains.

\subsubsection*{Virtual Knots}

\emph{Virtual knots} are an extension to the theory of classical knots~\cite{kauffman99SI} which classify all topological objects formed of ordered crossings, which generalises the theory of knot diagrams while keeping a sense of isotopy through Reidemeister moves. 
In particular, this includes those orderings which cannot be realised as plane projections of (closed) space curves in $\mathbb{R}^3$.  
They can be thought of as the objects represented by the set of \emph{all} Gauss codes, including sequences such as $1, -2, -1, 2$, which does not correspond to any closed knot diagram, as discussed above. 
In this sense, they provide a natural framework to describe \emph{open} diagrams, with endpoints that cannot directly be joined, so do not correspond to classical knots but have knot-like structure in their sequence of ordered crossings.

Many concepts from classical knot theory naturally generalise to virtual knots, such as the distinction between prime and composite virtual knots (including composites with classical and virtual components). 
Virtual knots are tabulated according to their minimum classical crossing number $n$~\cite{virtualknottableSI}, and they are denoted here as $vn_m$, following the tabulation of~\cite{virtualknottableSI}, as described in the main text.
The simplest nontrivial virtual knot, $v2_1$, has $n = 2$, and Gauss code $1,-2,-1,2$.
There are many more prime virtual knots for $n\geq 2$ than classical knots; complete tabulations only extend to virtual knots up to $n=5$. 
There are also up to three distinct chiral symmetric partners of a given virtual knot (compared to at most one partner of opposite chirality for classical knots): a mirror reflection of the diagram preserving the classical crossing signs, an inversion where all classical crossing signs are flipped, and the combination of both mirrors. 
As with the classical knots, we identify all chiral partners of the same virtual knot type as equivalent.

\cite{kauffman99SI} presents two further equivalent interpretations of virtual knots, both of which illustrate properties discussed in the main text. 
The first, convenient for diagrammatic representation, draws virtual knots as classical knot diagrams (without endpoints) but augmented with an additional crossing type at self intersection, the \emph{virtual crossing}, denoted by a circle around the intersection
(e.g. Fig.~2(b) of the main text). 
Virtual crossings do not have a sign and do not contribute to topological calculations, so the Gauss code follows only by considering the virtual diagram's classical crossings and ignoring virtual crossings entirely.  
In such virtual diagrams, virtual crossings can be manipulated by suitable generalisations of the classical Reidemeister moves, which can affect the configuration of virtual and classical crossings but do not change the virtual knot type; these moves are shown in Supplementary Fig.~1(b). 
In particular, virtual Reidemeister moves I and II can change the number of virtual crossings, and minimal virtual crossing number is an
invariant of virtual knots; those with minimum virtual crossing number zero are the classical knots, which make up a subset of the generalised, virtual knots. 
We describe a knot here as virtual if the minimum number of virtual crossings is greater than zero.

The other interpretation of virtual knots is as closed knot
diagrams drawn on surfaces with topology different to the standard
plane of projection (equivalent to its one-point compactification, the
2-sphere), i.e.~drawn on handlebodies with nonzero genus.  Any virtual
knot can be drawn as a knot diagram without virtual crossings on a
surface of sufficiently high genus~\cite{kauffman99SI}.  The virtual
crossings previously described are then interpreted as a consequence
of projection from the handlebody to a plane, in which case the
virtual crossings are intersections of two strands from different
bridges of the handlebody (likewise, a virtual knot diagram with
virtual crossings can be made a knot diagram on a handlebody by
replacing each virtual crossing with a handle which one strand passes
`along' and the other `under' the handle).  The minimum genus of any
handlebody on which the virtual knot can be drawn defines the
\emph{virtual genus} (hereafter referred to as the genus, although
this is distinctly different to the genus referred to in classical
knot theory~\cite{adams94SI}) of the virtual knot, and is therefore 0
for classical knots while any virtual knot must have genus at least 1.

Here, we are considering 2D open diagrams as virtual knots, and these interpretations of virtual knots relate directly to virtual closure of open diagrams (formed by projection of open 3D chains) considered in the main text. 
The virtual closure of an open knot diagram corresponds to adding a closing arc
between the open diagram's endpoints, where all intersections of this arc with the rest of the diagram make virtual crossings. 
All closure arcs are equivalent as they may be transformed to one another using virtual Reidemeister moves; the Gauss code only depends on the original open diagram, and does not change when the virtual crossings are altered. 
In fact, the virtual Reidemeister moves can be interpreted in terms of the endpoints of open diagrams, shown in Supplementary Fig.~1(c) in which the moves are effectively equivalent to different choices of closure. 

It is possible to consider the open diagram in these terms alone (i.e.~the open diagram is subject only to the three classical Reidemeister moves, but the endpoints are forbidden to pass over or under a strand creating (or removing) new crossings, Supplementary Fig.~1(d), otherwise the open diagram could be untangled to the trivial open curve); this would produce a \emph{classical knotoid}~\cite{turaev12SI}, a topological object that encodes information about the topology of the open curve, but whose classes are not isomorphic to the virtual knots~\cite{gugumcu16SI}. 
Representing knotoids by virtual knots loses some information -- for instance it may not be clear, from a virtual diagram, which arc at a virtual crossing is the virtual closure arc (i.e.~multiple, distinct knotoids give the same virtual knot).
However, in our analysis, we opt to work with virtual knots since their tabulation, invariants and other properties are a lot better developed and understood than for knotoids, and therefore are more convenient for application without new mathematics.
Only a small amount of information is apparently lost through the ambiguity of knotoids as virtual knots, which does not appear to
unduly limit topological analysis; this can be considered as a similar simplification to ignoring the chirality of knots.

Since all the virtual crossings resulting from virtual closure necessarily occur sequentially along the same arc, the genus of virtual knots obtained by closing open diagrams is at most one. 
That is, all the virtual crossings of the diagram may be removed by adding a single handle to the surface on which it is drawn, in between the endpoints of the open curve, and along which the closing arc runs.  
Not all genus one virtual knots can be represented in this way such that their virtual crossings occur sequentially; an example is shown in Supplementary Fig.~2(a), whose two virtual crossings can never be adjacent even under the application of (virtual) Reidemeister moves, although the knot can be drawn on a genus one surface sich as the planar diagram shown in Supplementary Fig.~2(b).  
The class of virtual knots that can be obtained from closures of open knot diagrams is therefore subset of genus one virtual knots, whose
minimal presentations pass around the torus exactly once in one generator direction, and at least once in the other.
This is related to the homology of the curve as drawn on a genus one handlebody: for any such diagram we can associate an index with the number of times a curve wraps around the torus in each direction, and for a virtual knot these homology indices must be of the form $(\pm1, j)$ for $|j| \geq 1$ (although this condition is not on its own sufficient due to the presence of more complex topologies with the same overall homology). 
We therefore refer to the virtual knots appearing as virtual closures of open curves as \emph{minimally genus one virtual knots}.

The virtual knots of genus one were studied and tabulated by~\cite{andreevna14SI}.  
Their description involves a virtual knot invariant that is a generalisation of the Kauffman bracket polynomial with two variables $a$ and $x$, calculated from the virtual knot diagram as drawn on the 2-torus. 
Each possible bracket smoothing of this diagram, $s$, is associated with a factor of $x^{\delta(s)}$, where $\delta(s)$ is the number of circles of nontrivial homology in a given smoothing. 
The polynomials for all minimally genus 1 virtual knots therefore have the form $x f(a)$, where $f(a)$ is a function of the knot which does not depend on $x$, and this property therefore allows all minimally genus one knots to be readily identified. 
The minimally genus one virtual knots of up to $n = 4$, in the genus one table~\cite{andreevna14SI} are, in the notation of that work: $2_1$, $3_1$, $4_1$, $4_2$, $4_3$, $4_6$, $4_7$, $4_8$ and $4_9$.  
In the complete virtual knot table~\cite{virtualknottableSI}, the diagrams which are explicitly minimally genus one are: $v2_1$, $v3_2$, $v4_{12}$, $v4_{43}$, $v4_{65}$, $v4_{94}$ and $v4_{100}$. 
After comparing knot invariants between the two tabulations, we were unable to find a partner in~\cite{andreevna14SI} for the minimally genus one $v4_{12}$ (i.e.~it appears to be an erroneous omission). 
Thus, from ~\cite{andreevna14SI} we could identify three further minimally genus one virtual knots than the complete table, with this property also confirmed via the Kauffman bracket method; these correspond to $v4_{36}$, $v4_{37}$ and $v4_{64}$ in~\cite{virtualknottableSI} (up to chiral mirrors). 
This relationship would be difficult to see by direct inspection of the diagram, and Supplementary Fig.~\ref{fig:supp_fig_3} demonstrates the equivalence of the different presentations for $v4_{64}$, via a combination of virtual Reidemeister moves and planar isotopies. 
All other minimally genus one examples agree in the two tables, and we believe that this completes the full set of minimally genus one virtual knots with up to four classical crossings.

Just as with classical knots, we identify virtual knot types by calculating virtual knot invariants (which are, in many cases, generalisations of classical invariants, such as the Kauffman bracket polynomial already discussed).  
Typically it is more computationally expensive to discriminate virtual knots than classical knots of the same minimum crossing number $n$. 
The basic procedure of invariant calculation is similar to that of classical knots, although now virtual crossings may also be algorithmically removed via virtual Reidemeister moves I and II. 
This does not directly affect the classical crossings, but may allow more of them to be removed. 
The Alexander polynomial has a number of extensions in virtual knot theory; we work with the two variable \emph{generalised Alexander polynomial} $\Delta_g(s,t)$~\cite{kauffman03SI}.  
As with classical knots, the calculation is significantly faster evaluated at constant values of $s$ and $t$, and we use the combinations ($s=-1$, $t=e^{2\pi i/3})$), ($s=-1$, $t=i$) and ($s=e^{2\pi i/3}$, $t=i$).
However, in contrast to classical knots, the generalised Alexander polynomial is not enough to distinguish the two simplest virtual knots possible from open curves, $v2_1$ and $v3_2$, as well as some other simple virtual knots (the next are $v4_{36}$ and $v4_{65}$, but although they are relatively simple these do not contribute significantly to any of our analysis).  
When necessary (but primarily in the case of $v2_1$ and $v3_2$), we resolve this ambiguity using the \emph{Jones polynomial} $V(q)$~\cite{jones85SI}, which is a classical knot invariant that extends to virtual knots without modification.
Since computation of the Jones polynomial takes exponential time in the number of crossings~\cite{adams94SI,knotatlasSI}, we compute it only at the constant $q=-1$ (sufficient to distinguish $v2_1$, $v3_2$, etc.), and only when our chosen values of $\Delta_g(s,t)$ are not sufficiently discriminatory to identify the virtual knot.

Virtual knot invariants for each of the virtual knots with up to four classical crossings can be found in the online knot table of~\cite{virtualknottableSI} or, for the Kauffman bracket variant explained above, in~\cite{andreevna14SI}. 
Supplementary Table~\ref{table:supp_tab_1} further shows the values of $\Delta_g$ and $V$ for each of the minimally genus one virtual knots in these tables, which together are clearly sufficient to distinguish all relevant knot types.

\end{document}